# Scaling Laws Governing Droplet Spreading and Merging Dynamics on Solid Surfaces: A Molecular Simulation Study

Ertiza Hossain Shopnil[a], Jahid Emon[b,*], Md. Nadeem Azad[c], and Dr. A.K.M. Monjur Morshed[d]

[a]*Department of Mechanical Engineering, Virginia Tech, Blacksburg, VA 24061, USA*
[b]*Department of Mechanical Science and Engineering, University of Illinois Urbana-Champaign, Urbana, IL 61801, USA*
[c]*Department of Aerospace and Mechanical Engineering, University of Notre Dame, Notre Dame, IN 46556, USA*
[d]*Department of Mechanical Engineering, Bangladesh University of Engineering and Technology (BUET), Dhaka, Bangladesh*

**Corresponding Author: jemon2@illinois.edu (Tel: 447-902-7621)*

## Abstract

This study employs molecular dynamics simulations to investigate droplet dynamics when a stationary droplet on a solid surface is struck by another droplet of similar size from above. The focus is on the jumping behavior of the merged droplet and the associated energy conversion. The process is primarily governed by the amount of energy converted into kinetic energy after dissipation. At high impact velocities, the energy conversion efficiency becomes constant, with only about 1% lost due to surface adhesion—an effect that diminishes with increasing velocity. Factors such as impact velocity, droplet size, surface texture, and wettability significantly influence the jumping velocity. Scaling laws are developed for the maximum spreading time, spreading factor, and restitution coefficient based on the Weber (We) and Reynolds (Re) numbers, which differ from those for single-droplet impacts. On superhydrophobic surfaces, the spreading time is approximated as $t_{sp} \approx 3r/V_i$, and its dimensionless form scales linearly with $We^{0.31}$. The general scaling law for the spreading factor is $\beta_{max} \sim We^{0.5\alpha}Re^{\alpha}$, where $\alpha = 0.1$ for velocity-dependent and $\alpha = 0.24$ for velocity-independent spreading regimes.

## 1. Introduction

Droplet collision, reshaping, and bouncing is a complicated process that occurs very often in both natural[1,2] and industrial processes.[3–6] When a single droplet strikes a surface, it undergoes a sequence of events: hitting, spreading, receding, and rebounding on that surface.[7–9] However, when a single droplet strikes another motionless droplet resting on the surface and then impacts the surface, it undergoes an additional step known as coalescence before impact.[10] This phenomenon has significant applications in nano-inkjet printing,[11–13] surface coating,[14] surface cleaning,[15–19] and spray cooling.[20] Anti-icing features in high altitudes can require insight into this phenomenon.[21–23] Additionally, tapping the renewable blue energy that is energy from natural water, such as using piezoelectric to harvest raindrop energy, demands an understanding of the mechanism and energy conversion of droplet impact.[24] Recently, Wu *et al.*[25] investigated the conversion of the kinetic energy of water droplets striking a charged surface into electrical energy. This study presents a novel approach to energy harvesting, making use of the droplet impact phenomena. So, this behavior has been the subject of many recent research, including both experimental,[26–31] and computational investigations.[32–38]

Two dimensionless numbers that help to describe this phenomenon are the Weber number, $We = \frac{\rho D_0 V_0^2}{\gamma}$ and Reynolds number, $Re = \frac{\rho D_0 V_0}{\mu}$. Density $(\rho)$, diameter $(D_0)$, velocity $(V_0)$, surface tension $(\gamma)$, and dynamic viscosity $(\mu)$ are the properties that determine these dimensionless numbers. The mechanism is explained with parameters such as the maximum spreading time, the maximum spreading factor, and the restitution coefficient. The duration necessary to attain maximum contact with the surface or maximum diameter $D_{\max}$ while spreading, is known as the maximum spreading time. Where the maximum spreading factor is $\beta_{\max} = \frac{D_{\max}}{D_0}$. The impact velocity, $V_0$ and the induced jumping velocity, $V_{\text{jump}}$ help to measure the restitution coefficient, $\epsilon = \frac{V_{\text{jump}}}{V_0}$. With the help of these parameters, this droplet collision has been investigated at the macroscopic level both experimentally[39–42] and numerically.[43–47] Koishi *et al.*[35] identified the deformation of droplets to be necessary for the bounce back of droplets on flat and nano-structured surfaces in nanoscale with the help of molecular dynamics. As the accuracy of macroscopic models drops in the case of nanoscale, Gao *et al*.[36] utilized molecular dynamics to develop a model for a single droplet impacting a rough surface. Following this work, Wang *et al*.[37] developed scaling laws for governing parameter maximum spreading factor for single droplet impact.

While numerous studies explain the phenomenon of a single droplet striking the surface, there is a lack of equivalent investigations concerning the collision that occurs when a moving droplet impacts a pre-existing immobile droplet on the surface. In reality, the latter phenomenon has more potential as it presents more likely circumstances. In practical applications, it is not uncommon for one droplet to collide with another droplet already present on the surface, possibly injected earlier. Therefore, there is a need to develop a quantitative study of the governing characteristics that are relevant to this phenomenon. This will facilitate the adoption of this phenomenon in innovative ideas and nanoscale devices, particularly for nano-injection applications.[11] Xu *et al*.[31] investigated the phenomenon by varying the temperature of the impacting droplet while keeping the temperature of the immobile droplet constant, aiming to determine the optimal temperature for the rebounding characteristics of the merged droplet.

However, in previous studies, no comprehensive analysis has been conducted for a droplet striking another immobile droplet on a rigid surface, and the effects of droplet size, surface roughness, and wettability on this phenomenon. This study presents a quantitative analysis at the nanoscale using molecular dynamics simulation, revealing insights into the energy conversion, droplet dynamics, and previously unexplored implications of droplet size, surface roughness, and wettability. A comparison of this phenomenon with traditional studies of a single droplet striking a surface is developed based on different factors. Through examining the interactions, modified scaling laws have been developed for important governing parameters like maximum spreading time, spreading factor, and restitution factor. These findings from the research provide a foundation for nano-injection technologies and the design of novel devices that utilize droplet collisions over surfaces.

## 2. Methodology

Molecular dynamics simulations were performed with the LAMMPS package[48] and the initial geometry was generated by Atomsk.[49] The simulation model with a flat surface initially is shown in Figure 1 (a). To evaluate the surface roughness effect, two types of structures were constructed on the flat surface of Figure 1 (a): grooves and nano-pillars (NP), shown in Figure 1 (b) and (c), respectively. The initial configuration consisted of a three-dimensional box with periodic boundary conditions applied to all the boundaries. Copper (Cu) has been chosen as the substrate, modeled with a 3.61 Å lattice constant as a face-centered cubic beneath the droplets. The substrate size was optimized to accommodate the highest spreading of the merged droplet after collision. The radius of the droplets ranged from 3 $nm$ to 7 $nm$. For the grooved surface, the height and thickness of each groove were 0.7 $nm$ and 1.2 $nm$ and the gaps between adjacent grooves were 3 $nm$. For the surface containing nano-pillars, the height and width of each nano-pillar were 0.7 $nm$ and 1.2 $nm$. Three times the droplet radius was chosen as the vertical distance between the droplets' centers to minimize simulation time and enable clear observation of the phenomenon.

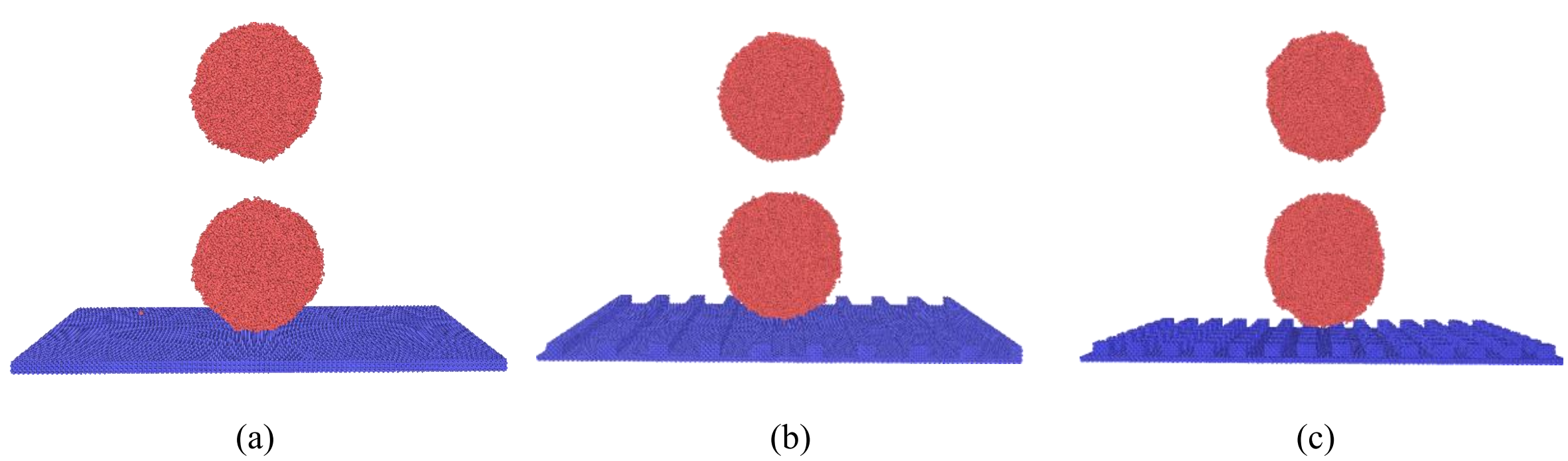

**Figure 1.** Initial configuration of the simulation domain, for the droplet on a (a) flat surface, (b) grooved surface, and (c) the surface containing nano-pillars (NP).

The coarse-grained water model, or mW model, was used in the simulation to simulate the water nanodroplets.[50] The mW properties of water needed for calculating the Ohnesorge number (Oh), Reynolds number (Re), and Weber number (We) were taken from a recent study.[51] The Stillinger-Weber potential was used to describe interactions between mW atoms in the mW model.[50] The mW density is 1000 $kg/m^3$, the mW viscosity is 310 $\mu Pa{\cdot}s$, and the mW surface tension is 65.4 $mN/m$.[51] The mW model considers oxygen and hydrogen as a single molecule to reproduce the nanoscale fluid dynamics of real water without the hydrogen atoms' reorientation. This is reflected in the mW model's viscosity, which is three times lower than that of actual water.[50,52] The mW model has previously been used in studies of droplet coalescence that highlight its applicability in the MD simulation of this study.[53,54] The Lennard-Jones potential was used to describe interactions between other molecules, and the size parameter was set to $\sigma = 3.92$ Å. The energy parameter was gradually adjusted from 0.05 to 0.01 $kcal/mol$ during equilibration to increase the hydrophobicity of the surface. The value of $\epsilon = 0.01$ was chosen to achieve a contact angle of approximately 180 degrees and ensure superhydrophobicity between the substrate and the water droplet. $\epsilon = 0.02$ and 0.03 were taken for contact angles $\sim 165$ and $\sim 155$ degrees respectively. The cutoff distance was set at 13.0 Å, and the timestep was adjusted based on the number of atoms in the system. Timesteps of 1 $fs$, 5 $fs$, and 10 $fs$ were simulated for several cases. The timestep of 5 $fs$ was selected for droplets up to 5 $nm$ in size, while a timestep of 10 $fs$ was chosen for droplets larger than 5 $nm$, as these provided greater accuracy while optimizing computational time.

During the equilibration stage, the substrate molecules were immobilized by setting their forces to zero. The entire system was relaxed using the conjugate gradient (CG) method and equilibrated with the Berendsen thermostat at 300 $K$, employing damping constants of 100 times the timestep for the thermostat. Subsequently, the system underwent a canonical ensemble, NVE (constant atom number, volume, and energy), with a relaxation period needed for each case to reach a stable total energy. All constraints were removed after equilibration, and the droplet at the top was given a downward velocity. The vertical component of the mass center's velocity was calculated by averaging the velocity per atom. This was defined as the jumping velocity induced after the impact. Similarly, the vertical component of the force was determined by averaging the force per atom. The magnitude of the force exerted by the droplet on the wall helped to determine the time when the droplet completely detached from the surface.

## 3. Results and Discussion

### 3.1. Model Validation

To validate the present simulation model, an impact test of a single droplet on a superhydrophobic surface having a contact angle of nearly 180 degrees was conducted. This choice was made because the droplet dynamics in the scenario, where a moving droplet impacts a stationary droplet on the surface, have not been explicitly studied before. The maximum spreading factor from the present study for a 14 $nm$ diameter droplet was calculated and compared with that of the previous study by Wang *et al.*[37] as presented in Figure 2. In both studies, the mW model was utilized to simulate water droplet behavior. Wang *et al.*[37] pointed out that for We number $1 < We < 34$ the maximum spreading factor of the droplet impacting on a superhydrophobic surface follows the following scaling law:

$$\beta_{max} \sim We^{\frac{1}{5}} \sim We^{0.2} \quad (1)$$

From the present simulation for a 14 \$nm\$ diameter droplet, the power fit slopes were found to be 0.2115.

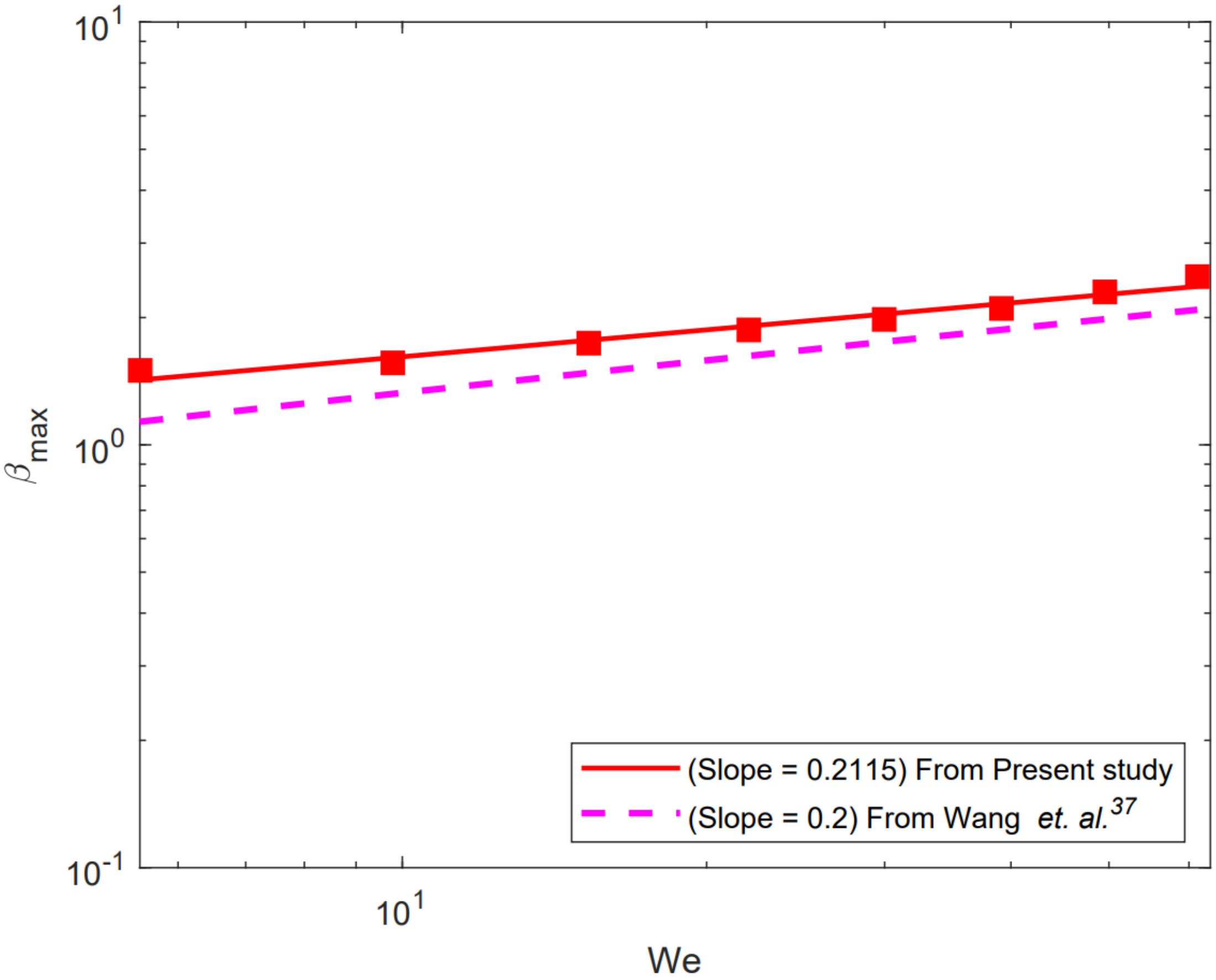

**Figure 2.** Maximum spreading factor as a function of We number from the present study and from Wang *et al*.[37] for 14 *nm* diameter droplet.

### 3.2 Jumping Process and Energy Conversion

The collision of a stationary droplet with a moving one causes the merged droplet to spread over the surface. An upward velocity is induced as a result of the reaction force of the impact and the reshaping process of the droplet into a round shape after spreading. The process is sequentially presented in Figure 4. Whether the induced velocity is enough to detach the merged droplet from the wall or not is determined by the available energy of the droplet after impact, which can be transformed into kinetic energy. In the case of a droplet impacting another stationary droplet, the sources of energy are the kinetic energy of the moving droplet and the excess surface energy released due to the merging of the individual droplets. During the spreading of the merged droplet, kinetic energy starts converting into surface energy, and at the maximum spreading, the majority of the kinetic energy transforms into surface energy. At this point, the merged droplet exhibits the highest surface area and surface energy. This surface energy is released as the droplet starts reshaping into a stable round shape, characterized by minimum surface area and surface energy. So, excess surface energy is available for conversion to kinetic energy, potentially causing jumping. However, a portion of the excess surface energy is dissipated in viscous dissipation during reshaping, where a portion of the kinetic energy of the moving droplet is lost in viscous dissipation while spreading to a maximum diameter. Loss of energy in viscous dissipation is always there, whether the droplet is spreading or reshaping. Also, the present study considers the droplet radius from 3 to 7 $nm$, and the Oh number is quite high (0.46 - 0.7), hence the loss in Viscous dissipation is likewise high. Xu *et al*.[31] highlighted that a portion of the excess energy is also lost by internal vibration due to the oscillation of the merged droplet. The oscillation of the droplet depends on the viscous effect. Zrnic *et al*.[55] studied nonlinear shape oscillations of Newtonian droplets, where the oscillation frequency decreases with the Oh number increase i.e. viscous effect causes a large dampening of the oscillating frequency and the damping rate rises with the Oh number. They found a cutoff of Oh 0.56 where the oscillation becomes zero. Also, while visualizing the simulation of the present work, the vibration due to the collision was hardly noticeable. So, there is no energy loss due to droplet oscillation here. When the surface energy fully converts to kinetic energy after the reshaping process from the spread form, the droplet does not leave the surface immediately. The maximum induced velocity is lowered by surface adhesion. So the energy to overcome surface adhesion comes from kinetic energy. To sum up, energy is wasted in viscous dissipation and wall adhesion.

The energy balance is:

$$E_{K(impact)} + E_{s1} + E_{s2} = E_{K(effective)} + E_{s12} + E_{vis} + E_{adh} \quad (2)$$

Here, $E_{K(impact)}$ is the kinetic energy of the moving droplet, $E_{s1} + E_{s2}$ is sum of the surface energies of two individual droplets before the collision, $E_{K(effective)}$ is the kinetic energy of the merged droplet while detaching from the surface, $E_{s12}$ is the surface energy of the merged droplet, $E_{adh}$ is the energy waste in overcoming adhesion of surface, and $E_{vis}$ is the energy lost in viscous dissipation.

Portion of total excess surface energy coming from the individual surface energies of the droplets after reshaping to a round shape:

$$E_{surf} = E_{s1} + E_{s2} - E_{s12} \quad (3)$$

Excess surface energy can be expressed as:

$$E_{surf} = 4\sigma\pi r^2 \left(2 - 2^{2/3}\right) \quad (4)$$

Where 'σ' is surface tension and '$r$' is the radius of the droplets.

The kinetic energy of the impact droplet and the merged droplet:

$$E_{K(impact)} = \frac{4}{6}\pi\rho r^3 V_i^2 \quad (5)$$

and

$$E_{K(effective)} = \frac{8}{6}\pi\rho r^3 V_j^2 \quad (6)$$

Where '$r$' is the radius of the impacting droplet, 'ρ' is density, $V_i$ is the impact droplet velocity and $V_j$ is the final velocity of the merged droplet while detaching from the wall.

Adhesive work:

$$E_{adh} = E_{K(induced)} - E_{K(effective)} \tag{7}$$

Where $E_{K(induced)}$ is due to the maximum velocity induced by the merged droplet and $E_{K(effective)}$ is due to the velocity at which the droplet is detaching from the wall.

Energy conversion efficiency:

$$\eta = \frac{E_{k(effective)}}{E_{K(impact)}} \tag{8}$$

Excess surface energy coming from individual droplets $E_{surf}$ serves as the dominant source of energy when the impact velocity is low ($ < $150 $m/s$) as shown in Figure 5 (a). The contribution of $E_{surf}$ is about 80\% when the impacting velocity is around 100 $m/s$ and it becomes negligible when the impact velocity is very high ($ > $700 $m/s$). At high impact velocities, the kinetic energy of the moving droplet contributes over 95\% to the total energy available for conversion.

When a single droplet impacts a surface, the energy source is only the kinetic energy of the impact droplet, because after spreading, the droplet will have the exact surface area it has initially as shown in Figure 3. No excess surface energy will come from the droplet's surface energy as the surface area is identical while impacting and during detaching from the wall, which is shown in Figure 3, as there is no merging of droplets. So, the term $E_{surf}$ is zero for the scenario of a single droplet impact. However, the energy dissipation is similar to the droplet impacting another droplet case, which is surface adhesion and viscous dissipation. Also, for single droplet impact, the expression of effective kinetic energy is:

$$E_{K(effective)} = \frac{4}{6}\pi\rho r^3 V_j^2 \tag{9}$$

Figure 5 (b) shows how the percentage of energy dissipation by surface adhesion is influenced by the moving droplet velocity for a droplet impacting another droplet. The percentage of energy lost to overcome surface adhesion linearly decreases with the increase in impacting velocity. At high impact velocity, the droplet has so much kinetic energy that it immediately detaches from the surface, disregarding the surface adhesion. Although the contribution to energy dissipation by surface adhesion is very small, approximately 1% at lower impact velocities, it is the crucial factor that terminates the droplet jumping when the impact droplet hits with lower kinetic energy

Figure 5 (c) illustrates the energy conversion efficiency with the impact velocity. It indicates the portion of surface energy after spreading, transformed into kinetic energy during the reshaping process. Energy conversion efficiency is low at lower impact velocity. It is because of the higher adhesive work at a lower impacting velocity.

Maximum efficiency is noticed between 200 - 300 $m/s$ impact velocity for all droplet sizes shown in Figure 5 (c). In this range, the droplet comes to a proper round shape after spreading and immediately before leaving the surface, as shown in Figure 4 (c). That's why all the surface energy is properly released just before the jumping is about to happen and converted into kinetic energy. Also, adhesive work is quite low in this range. At high impact velocity, the merged droplet leaves the surface with an unstable shape, i.e., before achieving a perfectly spherical shape as shown in Figure 4 (g), and not all the surface energy is utilized during the detachment of the merged droplet. But, this efficiency change is very small, and for the whole process, efficiency can be approximated as the constant value of efficiency achieved at high impact velocities. For instance, energy conversion efficiency for 6 $nm$ radius droplet impact is almost constant at about 4%. A small upward shift of the energy conversion efficiency curve is noticed in Figure 5 (c) with the increase in droplet size because, for smaller droplets, energy waste in viscous dissipation is more as the Oh number increases with the droplet size decrease (due to the increase of viscous effect over inertia and capillary effect).

Figure 5 (d) illustrates the energy conversion efficiency for the case of single droplet impact. The conversion efficiency is higher than that of the double droplet case at lower impact velocity. But, at higher impact velocity, the conversion efficiency for both cases of the droplet impact becomes very close. For a droplet impacting another similar-sized droplet, the droplet volume becomes double after impact and the spreading diameter is greater than that of a single droplet, thus the merged droplet has to work more against the adhesion of the wall while reshaping from the spread form, than the case of single droplet impact, for same impact velocity. Also, direct impact on a stiff surface gives a high reaction force to the response of the impact. On the other hand, when the moving droplet impacts another stationary droplet first and then the merged droplet hits the surface with less impact force, less velocity induction results, which is seen in Figure 6. That's why the energy conversion efficiency of a single droplet impact is about 10% whereas it is around 5% for a double droplet collision when the impact velocity is moderate. But, at high-impact velocities, the energy conversion efficiency of single droplet impact falls to 4%, as shown in Figure 5 (d). As discussed earlier for the case of double droplet collision, at high impact velocities, droplets do not assume a stable shape before detachment. In the case of a single droplet impact, the surface area of this unstable shape will be greater than that of its initial stage. Consequently, some of the energy becomes unavailable for conversion. Since the only source of energy is the kinetic energy of the moving droplet, the unavailability of energy due to the unstable shape significantly affects this process. This is the reason for the drastic drop in the energy conversion efficiency in Figure 5 (d). At low impact velocity, the single droplet has enough time to attain a stable spherical shape before detaching from the surface.

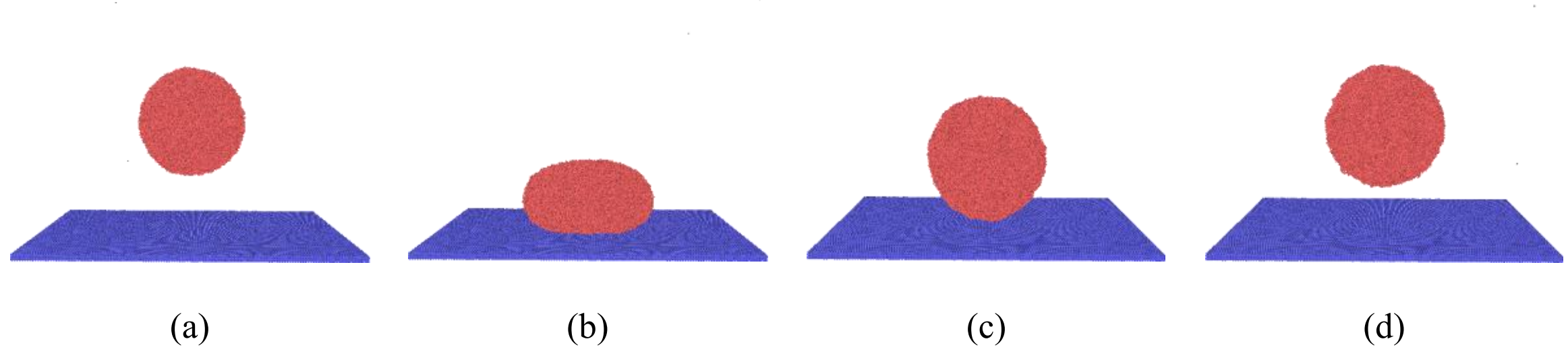

**Figure 3.** (a) A single droplet just before impacting the wall with a velocity of 100 m/s. The total energy of this stage is the kinetic energy and the surface energy of the droplet. (b) Spreading. Conversion of kinetic energy into surface energy of the spreading droplet. (c) Reshaping to a round stable shape. Releasing all the excess surface energy and the surface area is identical to the impact stage of the droplet. (d) Jumping off. A portion of the excess energy is converted into kinetic energy, which causes jumping. The rest of the energy is wasted in viscous dissipation and wall adhesion work.

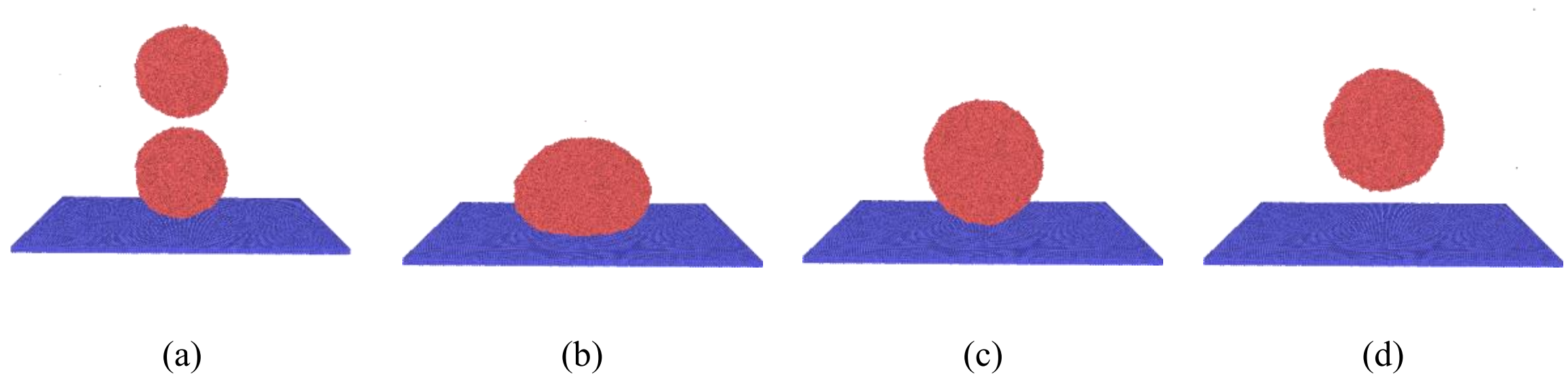

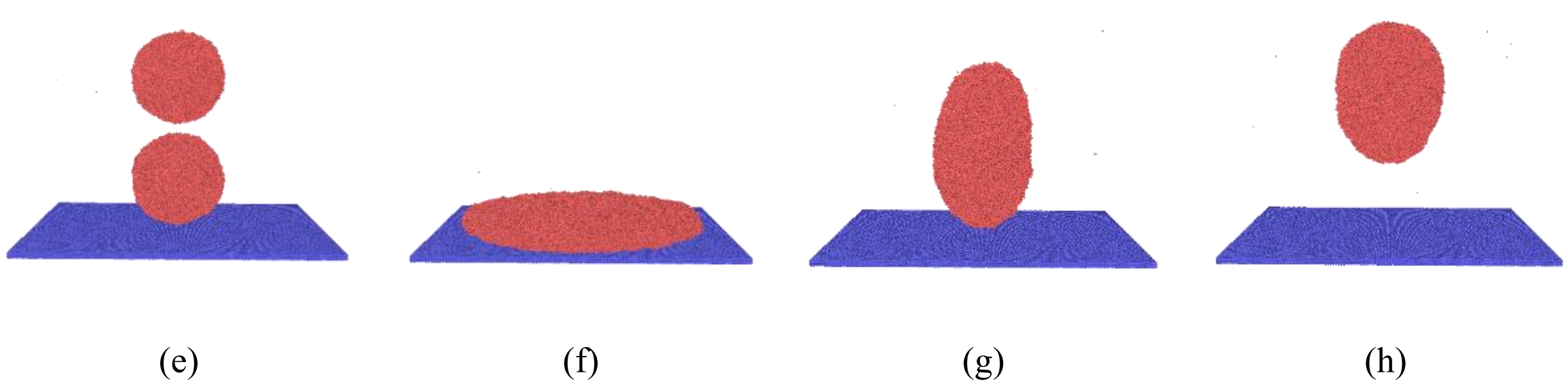

**Figure 4.** (a), (e) Just before impacting the stationary droplet with an impacting velocity of 100 m/s and 700 m/s respectively. The total energy of this stage is the kinetic energy of the moving droplet and the surface energy of individual droplets. (b), (f) Spreading, similar to the single droplet case. Reshaping to a (c) round stable shape, (g) unstable oval shape. The surface area of the merged droplet is less than the total surface area of the individual droplets. (d), (h) Similar to the single droplet shown in Figure 3.

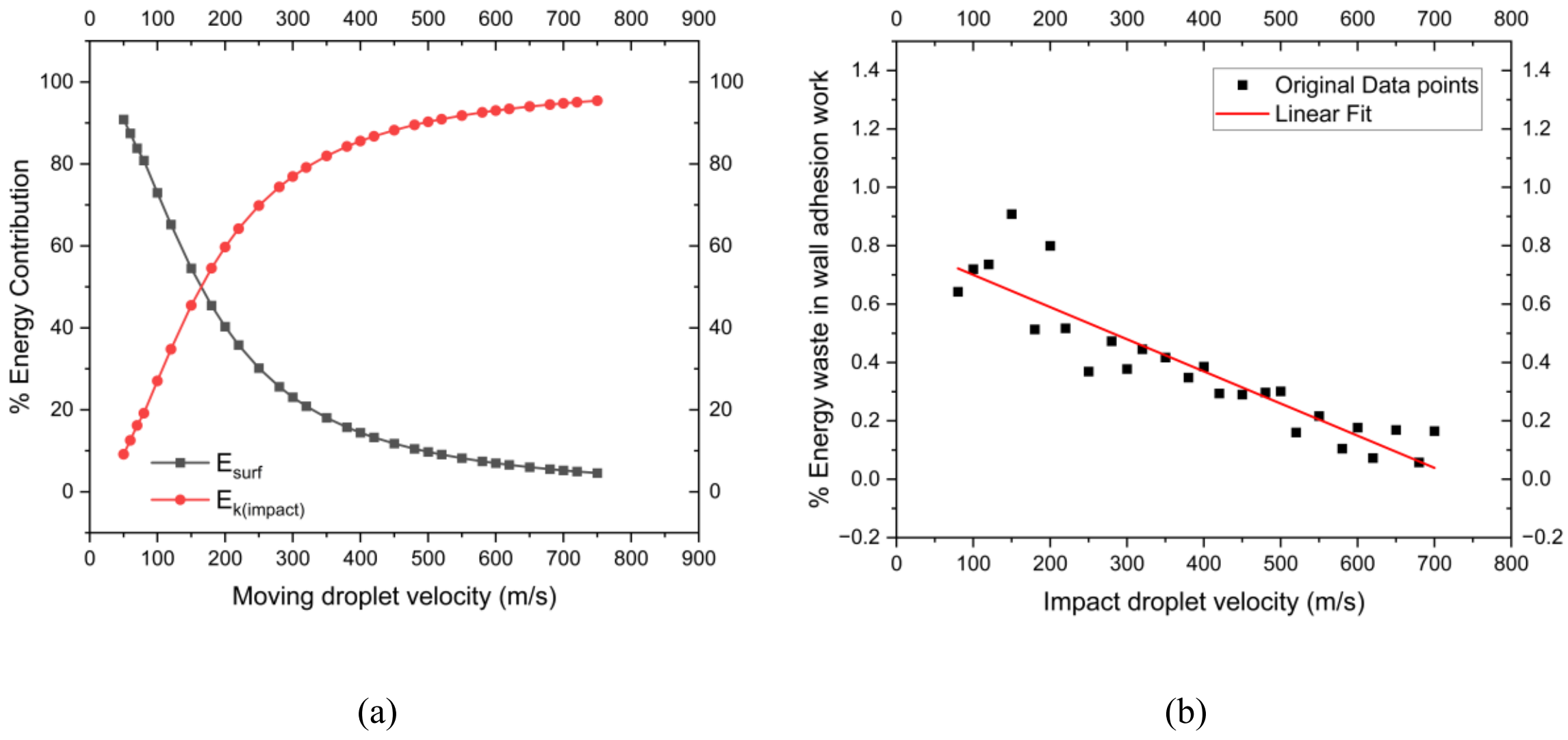

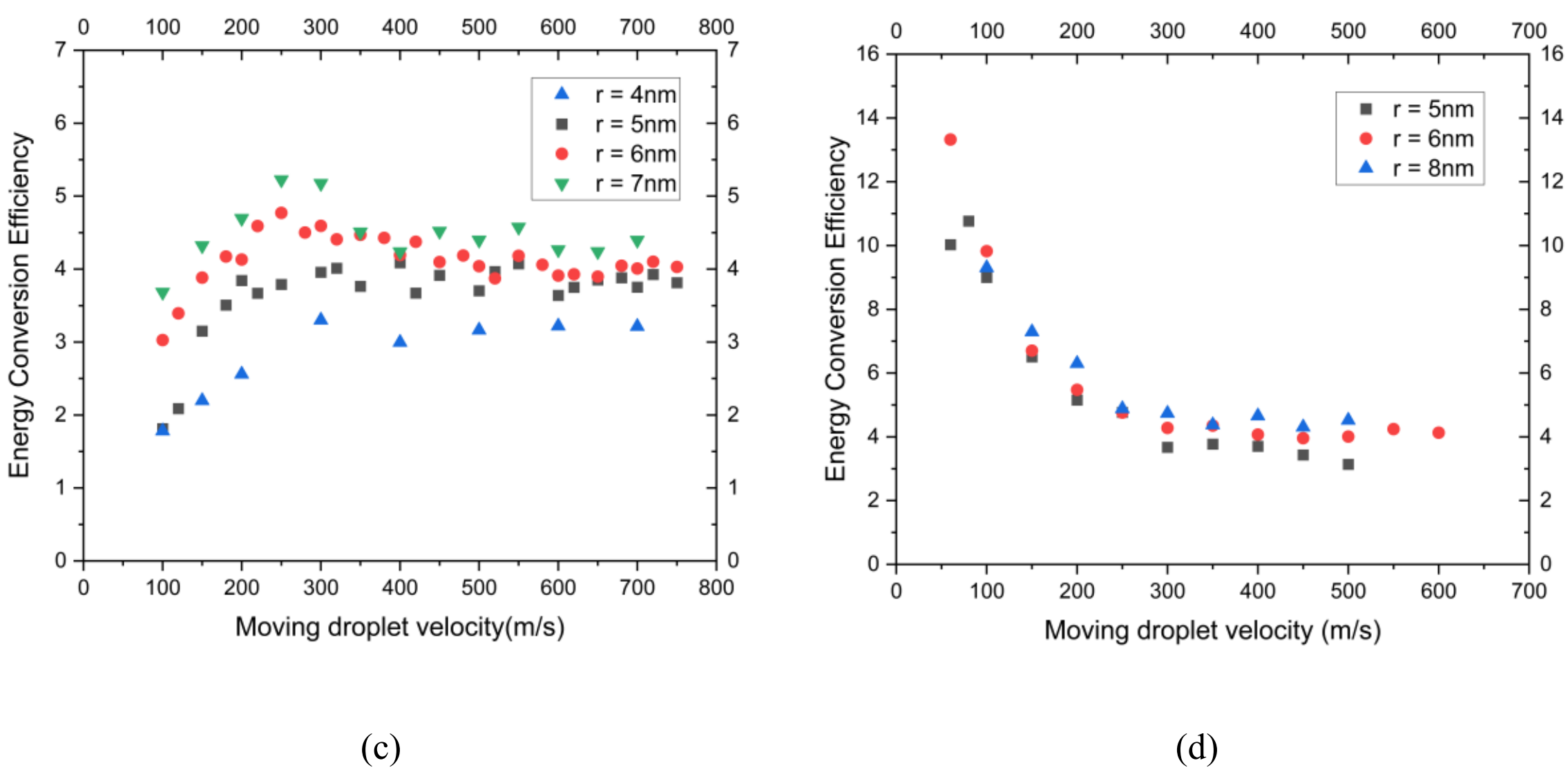

**Figure 5.** (a) For a droplet impacting another droplet, percentage contribution of the surface energy of individual droplets and kinetic energy of moving droplet, on total energy, (b) percentage waste of energy to overcome surface adhesion with droplet impact velocity for 6 *nm* droplet. (c) Percentage of total energy converted into effective kinetic energy, as a function of impact velocity for the droplet impacting another droplet, both having a radius of 4, 5, 6, and 7 *nm*. (d) The energy conversion efficiency for the single droplet impact having radius 5, 6, and 8 *nm*.

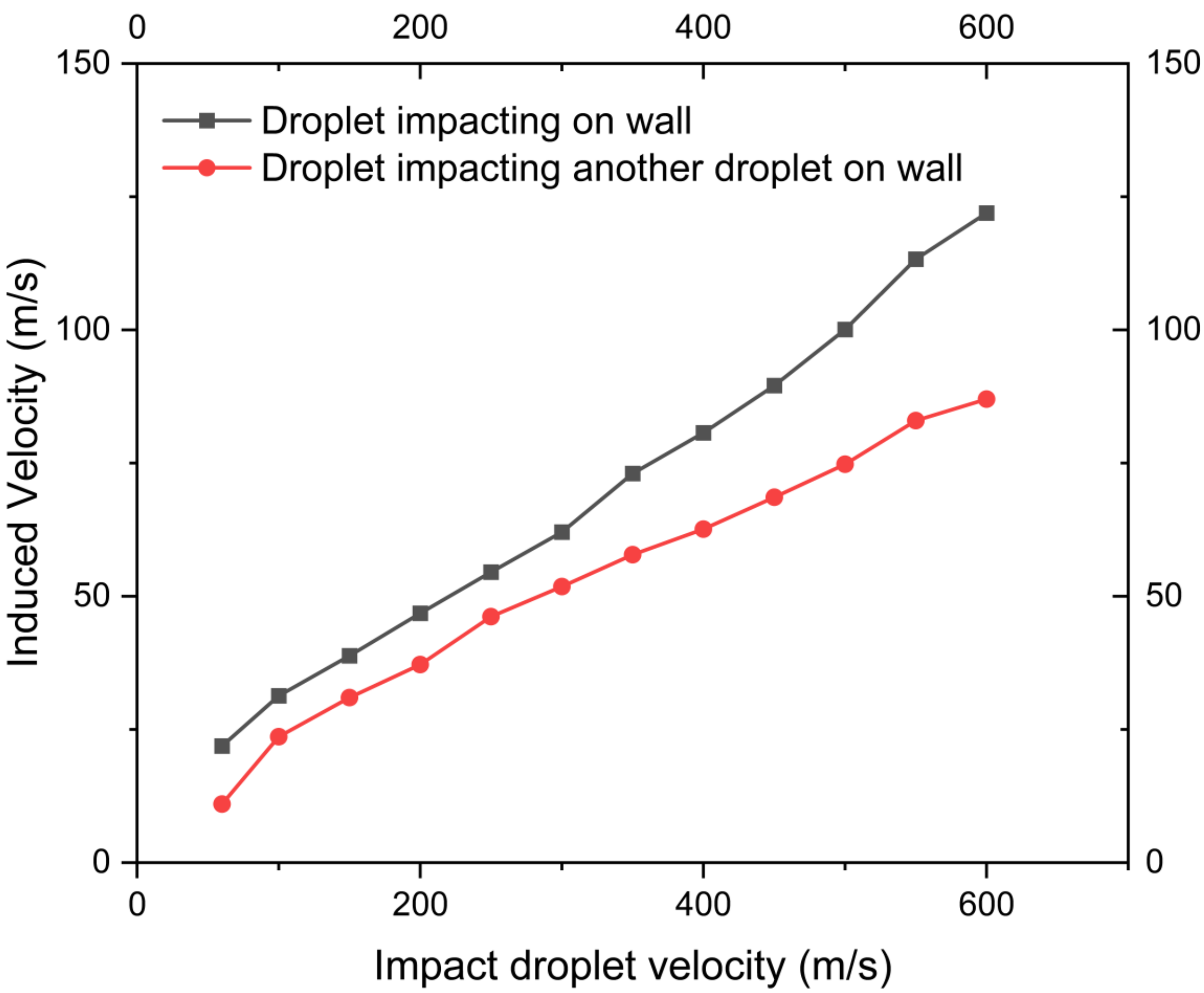

**Figure 6.** Comparison of the induced velocity as a function of impact velocity for a single droplet impacting directly on the wall and for a droplet impacting on another droplet on the wall. In both cases, the droplet radius is 6 *nm*.

### 3.3 Effects of Impact Velocity, Droplet Size, Surface Roughness, and Wettability

The induced velocity ($V_{ind}$) is a strong function of moving droplet velocity. The induced velocity is the velocity of the merged droplet after the collision. Figure 7 (a) illustrates that the induced velocity of the merged droplet varies linearly with the moving droplet velocity. Induced velocities of droplets having different sizes increase with the same slope of 0.13 with impact velocity. This phenomenon justifies that the conversion efficiency is nearly constant at a certain value as shown in Figure 5 (c) at moderate to high-impact velocities. Because, at impact velocities over 200 $m/s$, the kinetic energy of the moving droplet dominates and almost single-handedly controls the whole process. The velocity of the moving droplet is expressed in terms of a dimensionless number called We number in Figure 7 (b) and how induced velocity increases with We number is shown. Here the We number is determined only considering the moving droplet because the droplet situated on the wall has no velocity initially. The curve of Figure 7 (b) is plotted in a log-log graph and all five data sets are fitted by the power-fitted curve where the slope of the curve is 0.3894. So, the induced velocity of the merged droplet follows the following relation with the We number:

$$V_{ind} \sim We^{0.3894} \tag{10}$$

Additionally, Figure 7 (a) illustrates a larger induced velocity for larger droplet sizes at any impact velocity. Because, the energy conversion efficiency is relatively low for smaller droplets due to greater viscous dissipation, as discussed in the previous section.

The cessation of induced velocity occurs at low-impact velocities for two reasons. One obvious reason is the low kinetic energy of the impact droplet, which is insufficient to overcome viscous dissipation and adhesion forces. Another reason is the detachment of the stationary droplet from the wall just before the collision, which is caused by the attraction of the moving droplet approaching the stationary one. This phenomenon is observed for droplets making contact angles of approximately 180 degrees with the surface, indicating a pure superhydrophobic surface. In such cases, the merging occurs at a distance above the wall, and the merged droplet strikes the wall with shallow velocity, preventing it from jumping after impact. Consequently, the deformation or spreading of the merged droplet is minimal. However, at higher impact velocities, the stationary droplet does not have sufficient time to leave the surface before the collision. Nonetheless, the attraction of the moving droplet toward the stationary one is always there.

The surface roughness also influences the jumping dynamics similar to the single droplet impact as highlighted in a previous study.[36] Figure 8 depicts how velocity curves shift upward with the increase in surface roughness by creating grooves and nano-pillars on the surface. Vahabi *et al*.[56] showed the conversion efficiency from excess surface energy to kinetic energy can be increased by around 500% by creating a ridge on the surface in case of coalescence-induced jumping. This occurs due to the redirection of the velocity field inside the droplet. It is also applicable in droplet impact, and this phenomenon justifies the increase in velocity observed in the case of droplet impact after creating some nanostructures on the surface. Figure 9 illustrates how induced velocity increases with the increase in the We number on a surface containing nano-pillars when the droplet radius is 5 $nm$. The fitted curve gives a slope of 0.3420 whereas for the flat surface, it is 0.3894. So, the slope for the impact on a droplet situated on a flat surface is steeper than that of a rough surface. It is because, at lower impact velocity, the flat surface greatly affects the jumping process due to surface adhesion, the influence of which declines with impact velocity increase. However, in the case of rough surfaces, the adhesion effect is already low even for lower impact velocity. This is because, on the rough surfaces the droplets have points where it doesn't make contact with the surface. Therefore, for a droplet impacting a rough surface, the effect of surface adhesion increase or decrease doesn't affect the induced velocity much. This is another reason why the induced velocity curve shifts upward when the surface has roughness, as shown in Figure 8. That's why the slope for the flat surface is steeper, but the difference in slope is small and the scaling law (10) can also be applied to the droplet impact on the rough surface.

The jumping velocity of the merged droplet is also influenced by surface wettability. It is easier to detach the droplet from a surface with higher hydrophobicity. Figure 10 shows how the induced velocity for a 5 $nm$ radius droplet responds as surface wettability changes, for smooth and rough surfaces. In the case of a flat surface, induced velocity is strongly affected by the change in contact angle (CA) between the wall and the droplet as shown in Figure 10 (a). When the CA is close to 180 degrees, indicating a pure superhydrophobic surface, jumping occurs when the impact velocity is above 50 $m/s$. However, when the CA is reduced to near $\sim$165 and $\sim$155 degrees, to make the droplet jump from the surface, the impact droplet must have a velocity over 150 and 280 $m/s$, respectively. In contrast, the velocity curves of Figure 10 (b) and (c) are very close to each other for different contact angles (CAs). CAs shown in Figure 10 (b) and (c) indicate the CA a droplet would make if the surface were flat. As mentioned previously, surface roughness causes the droplet to make contact with fewer points on the surface, thereby increasing the hydrophobicity of the surface. That's why a flat surface where the droplet is making, for example, a contact angle of approximately 155 degrees, will exhibit a higher contact angle if nanostructures are constructed on the same surface.

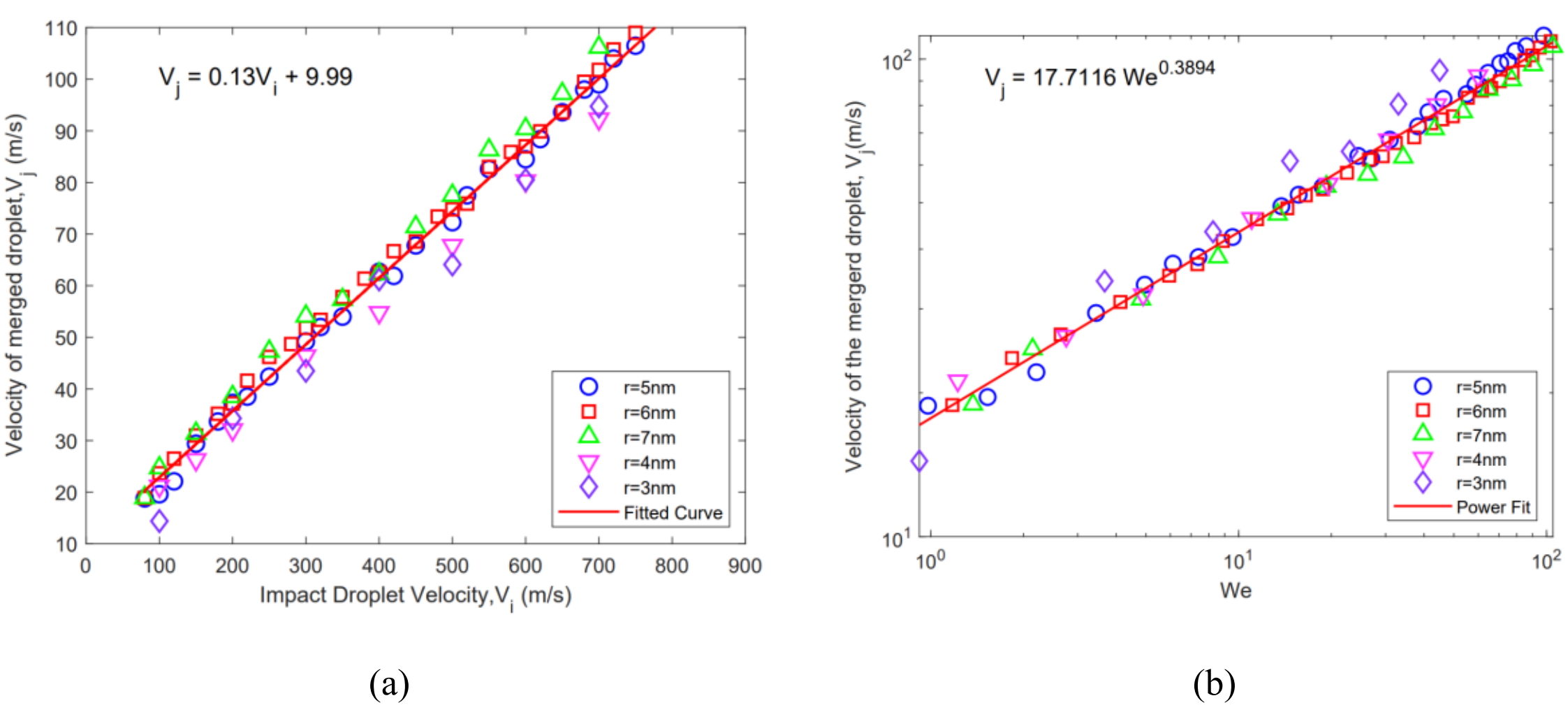

**Figure 7.** Induced velocity of the merged droplet as a function of (a) impacting velocity and (b) We number.

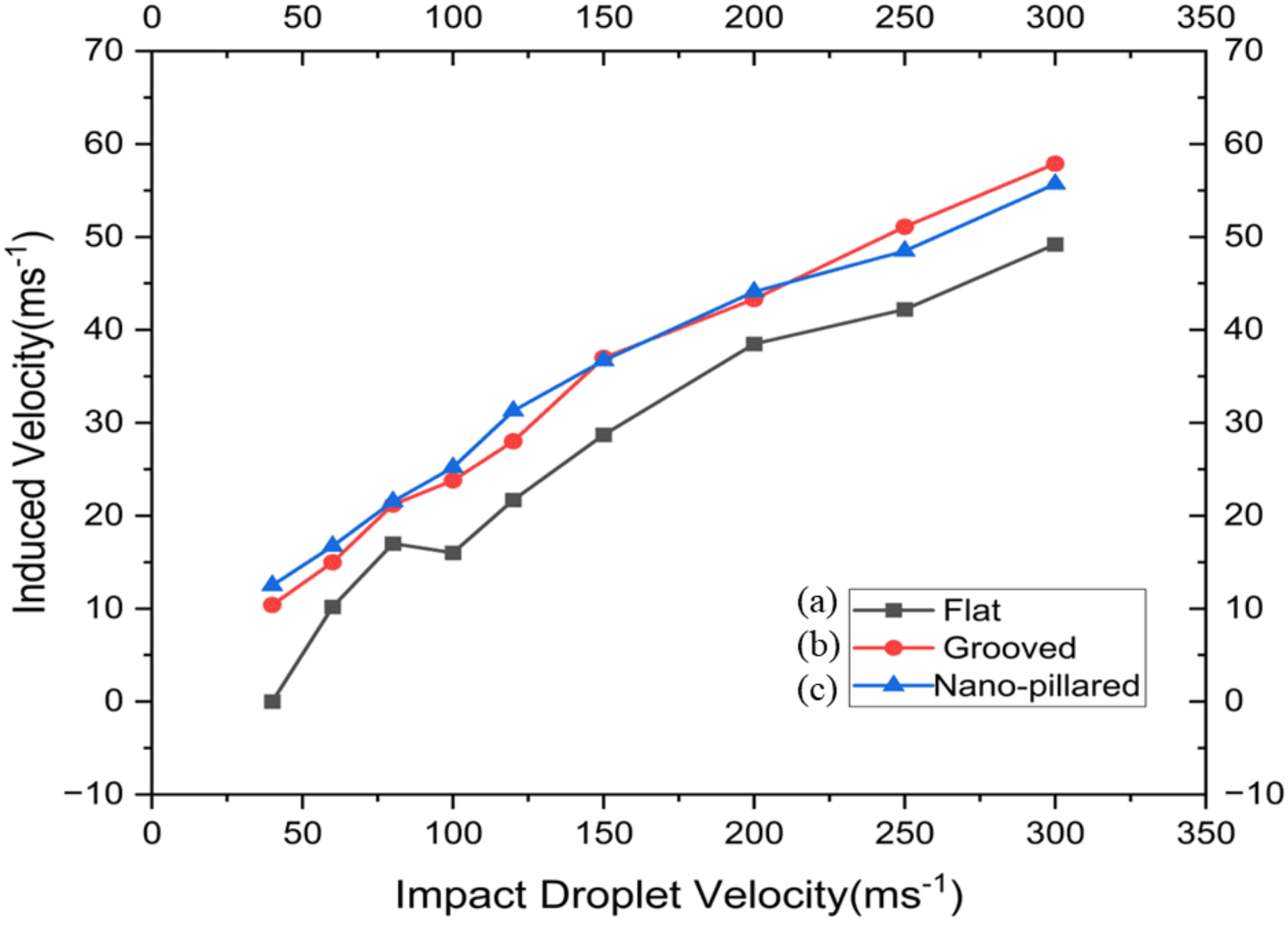

**Figure 8.** Variation in induced velocity on differently structured surfaces: a) flat, b) grooved, and c) nano-pillared surface.

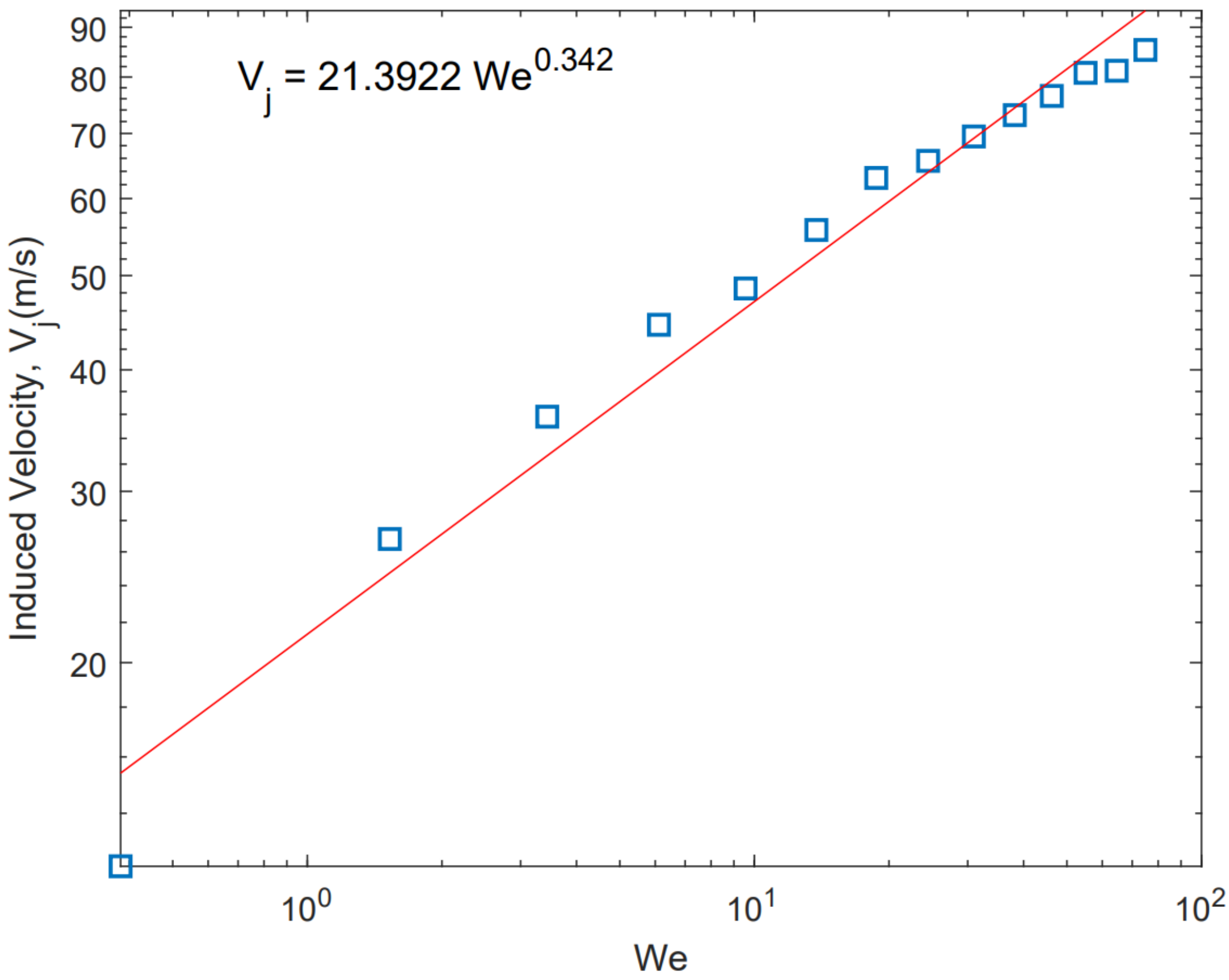

**Figure 9.** Induced velocity as a function of We number for droplet radius 5 *nm* and the stationary droplet is on a surface containing nano-pillars.

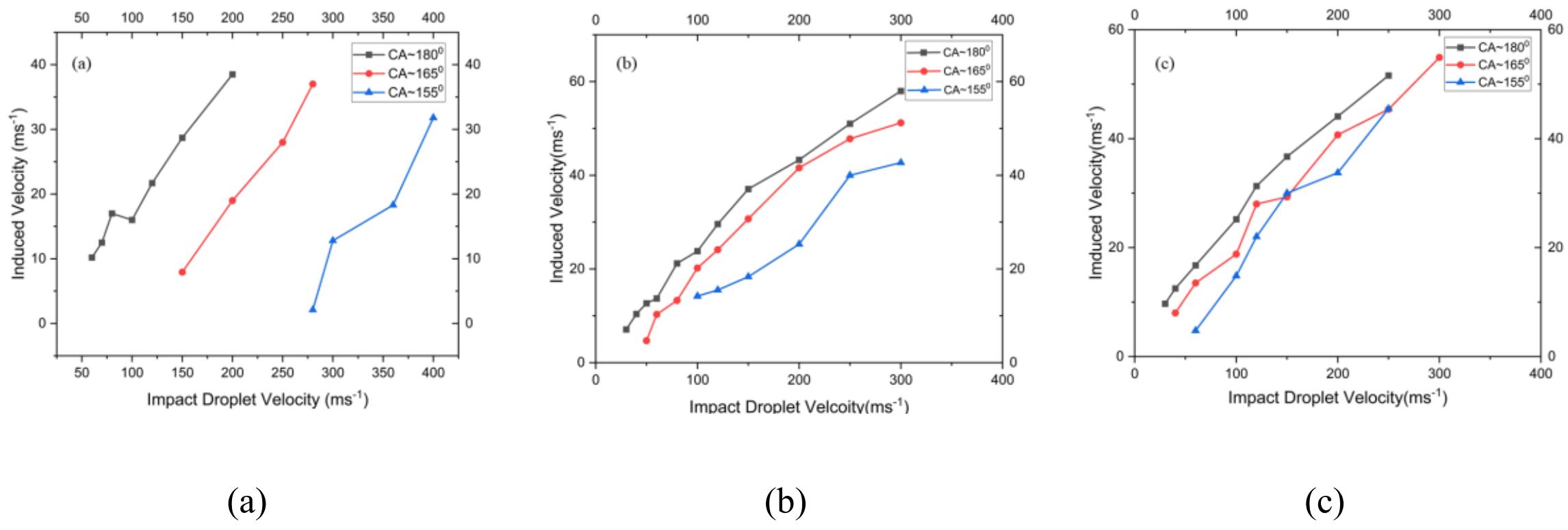

(a) (b) (c)

**Figure 10.** Variation in induced velocity on different structured surfaces: a) flat, b) grooved, and c) nano-pillared having different wettability.

### 3.4 Modified Scaling Law

#### 3.4.1 Maximum Spreading Time

After impact, the merged droplet expands across the surface until it reaches a maximum diameter. For single droplet impact, this spreading time is independent of impact velocity and it can be expressed as the following scaling law[37]: $t_{sp} \sim (D_o/V_o)We^{2/5}$. Gao *et al.*[36] showed spreading time reaches a constant value at a high impact velocity and they also found a similar relation as above. In this study of a stationary droplet on surface being impacted by another droplet, these relations will change. For a single droplet impact, the spreading time denotes the duration of the droplet remaining in contact with the surface from the moment of impact until it detaches. But in this context, one droplet is already on the surface, the spreading time is considered as the interval starting from when the moving droplet comes into contact with the stationary droplet until the entire merged droplet detaches from the surface. Figure 11 illustrates how spreading time varies with the velocity of the moving droplet. Comparable to the scenario of a single droplet impact, spreading time is no longer dependent on impact velocity at high velocities in this scenario as well. According to Figure 11, the velocity at which the spreading time becomes independent of the velocity of the impacting droplet is approximately 400 $m/s$. Additionally, all four cases (5 $nm$, 6 $nm$, and 7 $nm$ radius droplets on a flat surface, and the 5 $nm$ droplets on a nano-pillared surface) depicted in Figure 11 exhibit the same trend. Equations of the power-fitted curve for individual data sets are also provided in Figure 11. So, the droplets follow the following relationship between spreading time and impact velocity, for the present case of droplet impact:

$$t_{sp} \approx 3rV_i^{-0.32} \tag{11}$$

Where r is the radius of the droplet and $V_i$ is the impacting velocity of the moving droplet. In the case of droplet impact on a rough surface, the spreading time does not exactly follow the equation (11) which is depicted in Figure 11 for the 5 $nm$ droplet on nano-pillars, but the behavior is quite similar.

Figure 12 illustrates how normalized spreading time varies with the We number. The region of the We number where the spreading time is constant, is disregarded in Figure 12. Maximum spreading time, $t^*$ is normalized in the following way:

$$t^* = \frac{V_i}{D} t_{sp} \quad (12)$$

Where D is the diameter of the droplets before impact and $V_i$ is the velocity of the moving droplet. $t^*$ varies similarly with the We number for the droplets of all sizes shown in Figure 12. So, the modified scaling law is:

$$t^* \sim We^{0.31} \quad (13)$$

**3.4.2 Spreading Factor**

Another important parameter that characterizes the dynamics of impact droplet is the spreading factor which is the ratio of the initial diameter of the moving droplet to the maximum diameter achieved by the droplet after spreading on the surface $\beta_{max} = D_{max}/D_o$.

The correlation for single droplet impact, given by Wang *et al.*[37] for how $\beta_{max}$ varies with We and Re number ($\beta_{max} \sim We^{1/5}$ and $\beta_{max} \sim We^{2/3}.Re^{-1/3}$) will change in the present case of droplet impact onto another stationary droplet. For this study, the initial droplet diameter is taken as the diameter of the single droplet and the maximum spreading diameter is the diameter of the merged droplet after spreading.

As seen in the previous subsection, the spreading time is not influenced by the impact velocity for a range of We numbers and is dependent on impact velocity for another range. Also, it is seen in Figure 5 (c) that energy conversion efficiency becomes constant after an impact velocity range. Considering these factors, the behavior of the spreading factor is observed in two regimes of We number: high and low We number regimes. Separate relations of spreading factor with We and Re are found depending on We number regimes. The power-fitted curves of Figure 13 illustrate how the spreading factor changes with the We number and Re number. At low We number the $\beta_{max}$ varies with We and Re linearly in the log-log graph, making a slope of 0.1 and 0.19 approximately, as depicted in Figure 13 (a) and (b). So, $\beta_{max}$ follows scaling law $\beta_{max} \sim We^{0.1}$ at a low We number regime if only the We number is considered. Similarly, at a high We number regime, $\beta_{max}$ varies with $We$ and $Re$ making slopes 0.24 and 0.45 respectively, as shown in Figure 13 (c) and (d). The power of the Re number is found to be nearly double the power of the We number in both high and low We number regimes.

If the spreading factor is plotted against both We and Re numbers together using the slopes of We and Re numbers obtained from Figure 13, meaning if $\beta_{max}$ is plotted in a log-log graph against $We^{0.1}Re^{0.19}$ and $We^{0.24}Re^{0.45}$ for low and high We number regimes respectively, then the fitted curves give the slope of 0.5 for both high and low regimes of We number, as illustrated in Figure 14. Therefore, if both We and Re numbers are considered, then the scaling law for the spreading factor for the low We number regime is:

$$\beta_{max} \sim (We^{0.1} Re^{0.19})^{\frac{1}{2}} \tag{14}$$

Where 0.1 and 0.19 are the slopes of the $\beta_{max}$ vs We and Re number curves respectively, presented in Figure 13 (a) and (b).

For high We number:

$$\beta_{max} \sim (We^{0.24} Re^{0.45})^{\frac{1}{2}} \tag{15}$$

Similarly, where 0.24 and 0.45 represent the slopes of the $\beta_{max}$ vs We and Re number curves respectively, shown in Figure 13 (c) and (d). Hence, a general modified scaling law for the case of a droplet impacting upon a stationary droplet on a superhydrophobic surface is:

$$\beta_{\max} \sim (We^{\alpha} \cdot Re^{2\alpha})^{\frac{1}{2}} \sim We^{0.5\alpha} \cdot Re^{\alpha} \tag{16}$$

Where α represents the slope of the power fitted curves of $\beta_{max}$ vs We number. α = 0.1 and 0.24 for low and high We number regimes respectively for a droplet impacting another droplet on a solid surface.

### 3.4.3 Restitution Coefficient

The restitution Coefficient represents the ratio of induced velocity to impact velocity, $\epsilon = \frac{V_j}{V_i}$, where $V_j$ and $V_i$ are the jumping velocity of the merged droplet and impacting velocity of the moving droplet. The induced velocity of the merged droplet here is less than the velocity by which a single droplet departs from the surface after the impact. The reasons are described in the previous section. A study by Gao *et al*.[36] gives a relation between the restitution coefficient ( $\epsilon$ ) and the We number for a single droplet, which is $\sim We^{-0.341}$. To modify this scaling law of restitution coefficient for the present case of a moving droplet impacting a stationary droplet, how the restitution coefficient varies with the We number is presented in Figure 15, for droplets having radii 4, 5, 6, and 7 $nm$. $\epsilon$ follows the same trend with the We number regardless of droplet size and becomes constant at high We numbers. Hence, the power-fitted curve gives the following scaling law:

$$\epsilon \sim We^{-0.106} \tag{17}$$

As mentioned in previous sections, in the high We number regime, the kinetic energy of the moving droplet is the most dominant source of energy, and energy loss in adhesion is negligible, thus only loss of energy is in viscous dissipation. Therefore, only the kinetic energy of the droplet in motion affects the process, and energy conversion efficiency is also constant at a high We number regime. Thus, the ratio of induced to impact velocity becomes almost constant at high impact velocities as shown in Figure 15, meaning the merged droplet induces velocity similarly to the corresponding impacting velocity of the moving droplet. Though the trend of $\epsilon$ is similar for all droplet sizes shown in Figure 15, the value where $\epsilon$ becomes almost constant varies with the droplet size. For instance, for high We numbers, a moving droplet having a radius of $6nm$ will induce 15% of its impact velocity into the merged droplet after impact, whereas this value will be 13% for the 4 $nm$ droplet impact.

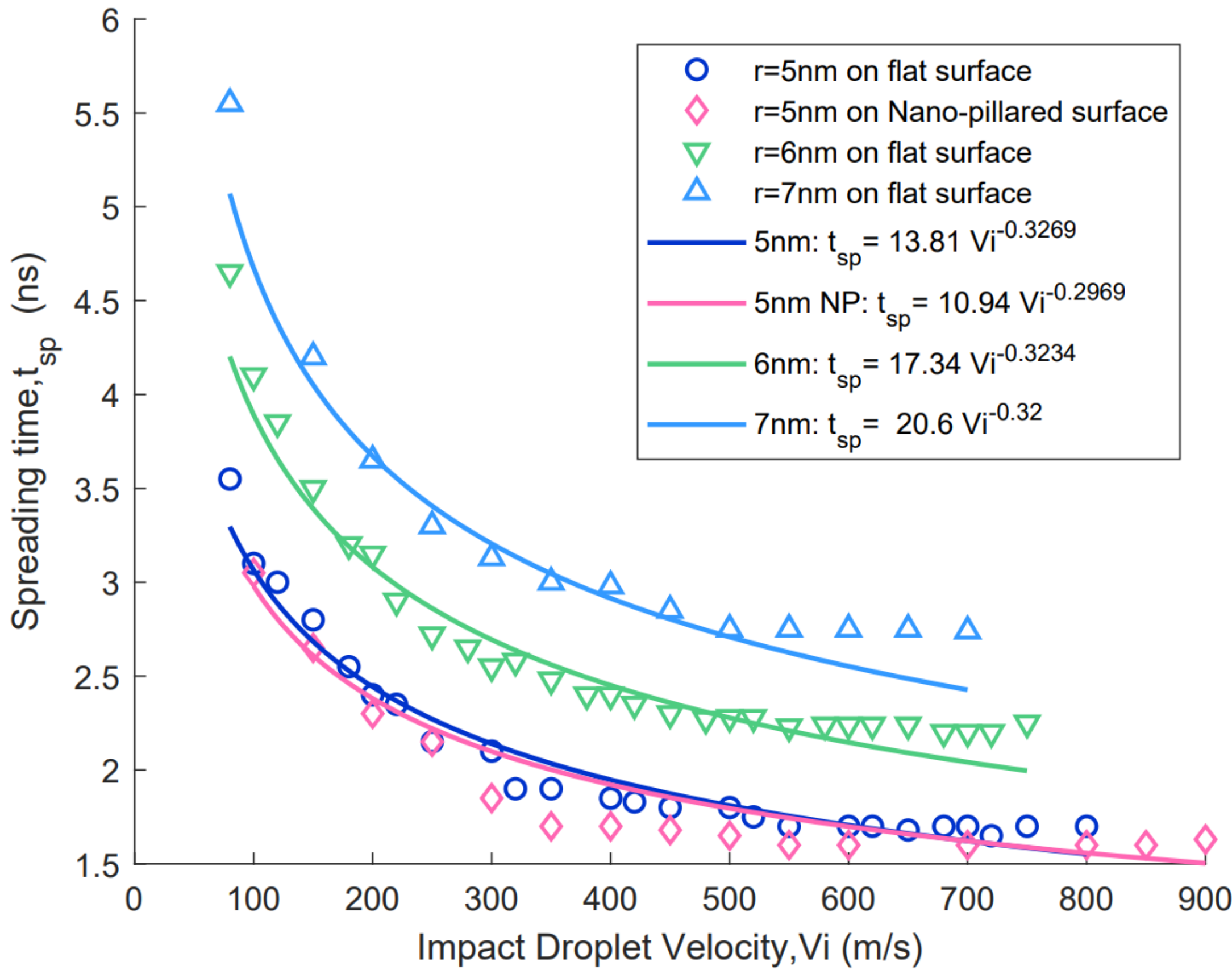

**Figure 11.** Variation of spreading time as a function of impacting velocity for droplets having radius 5, 6, and 7 *nm* on the flat surface and 5 *nm* droplet on a surface containing nano-pillars (NP).

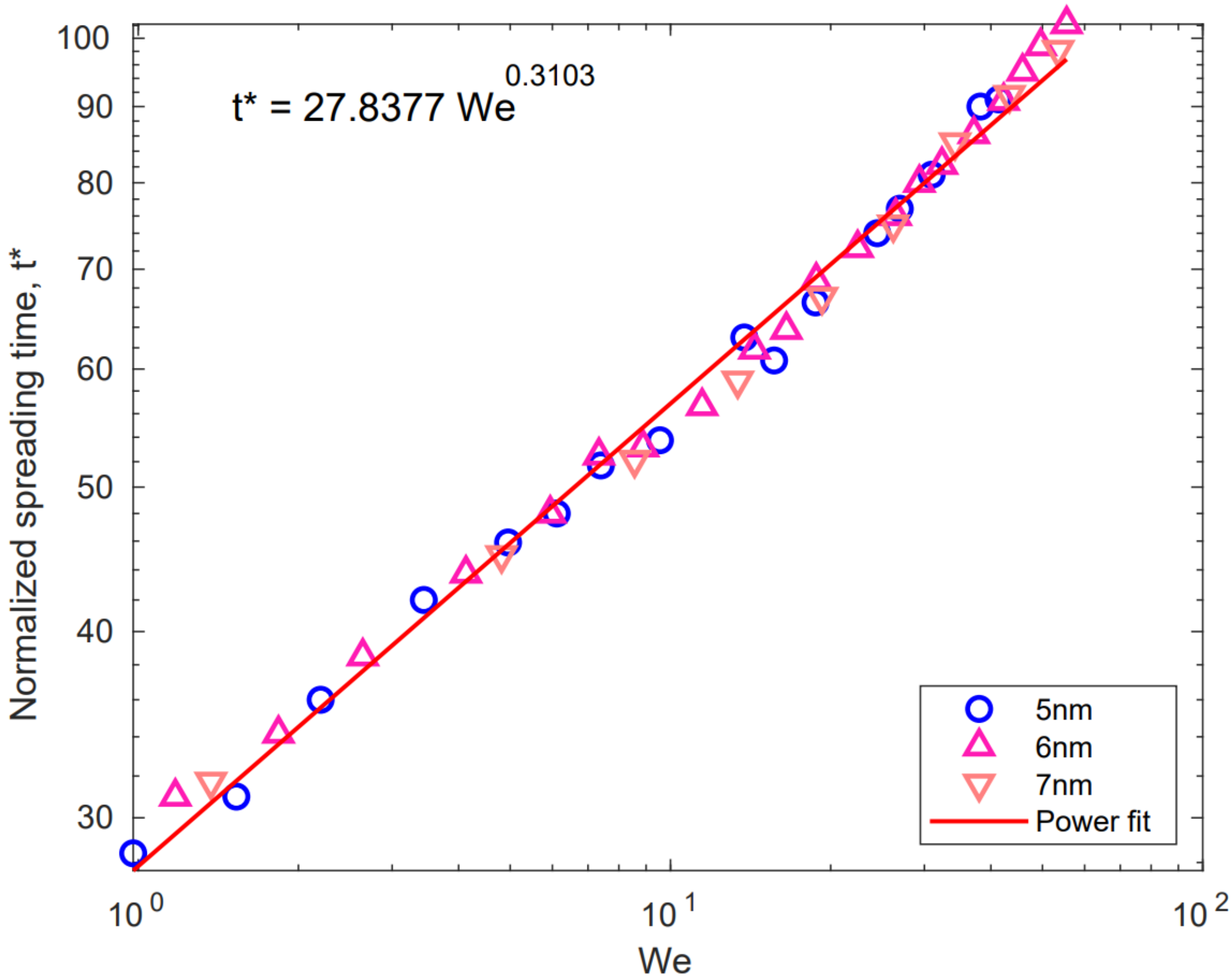

**Figure 12.** Variation of normalized spreading time as a function of We number. Data points are fitted for 5, 6, and 7 *nm* droplets on a flat superhydrophobic surface (CA ~ 180).

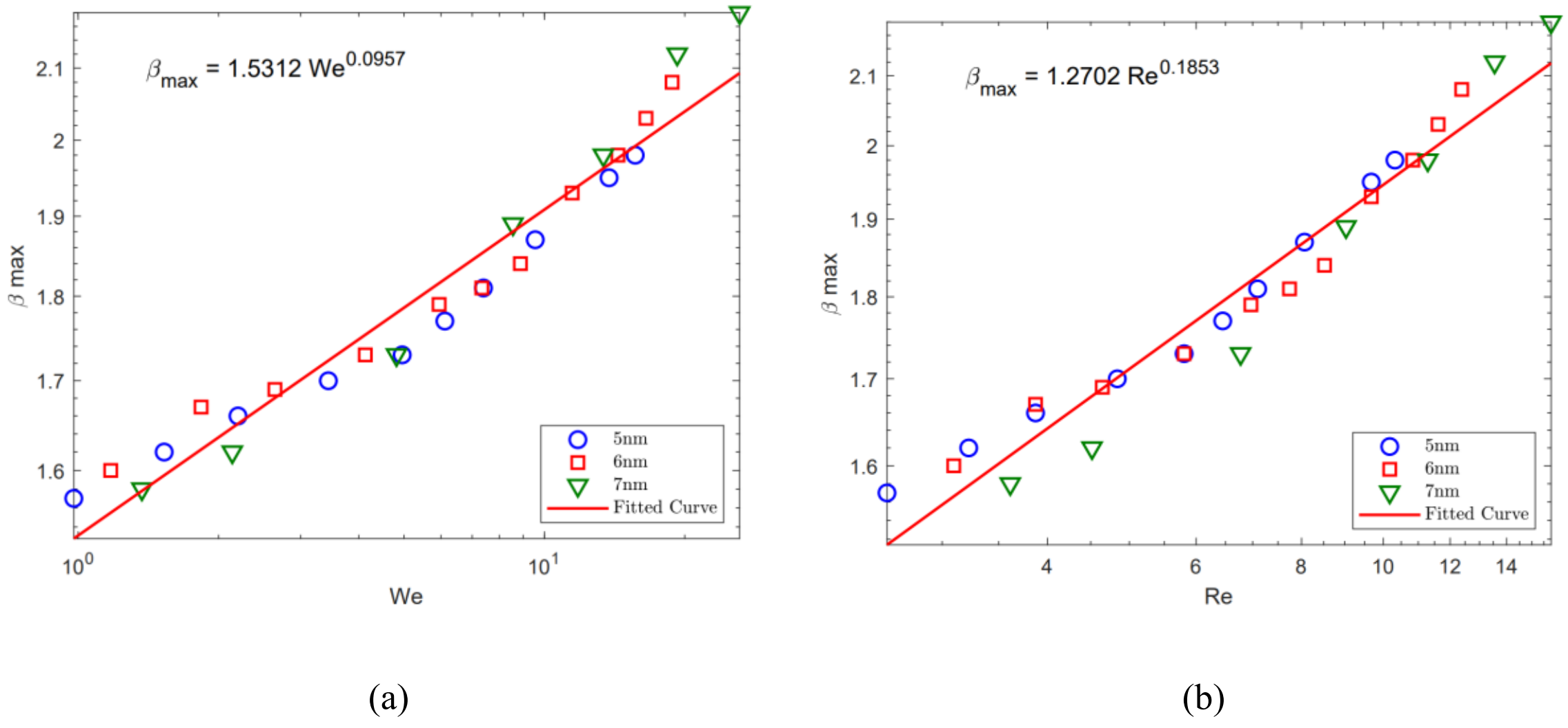

(a) (b)

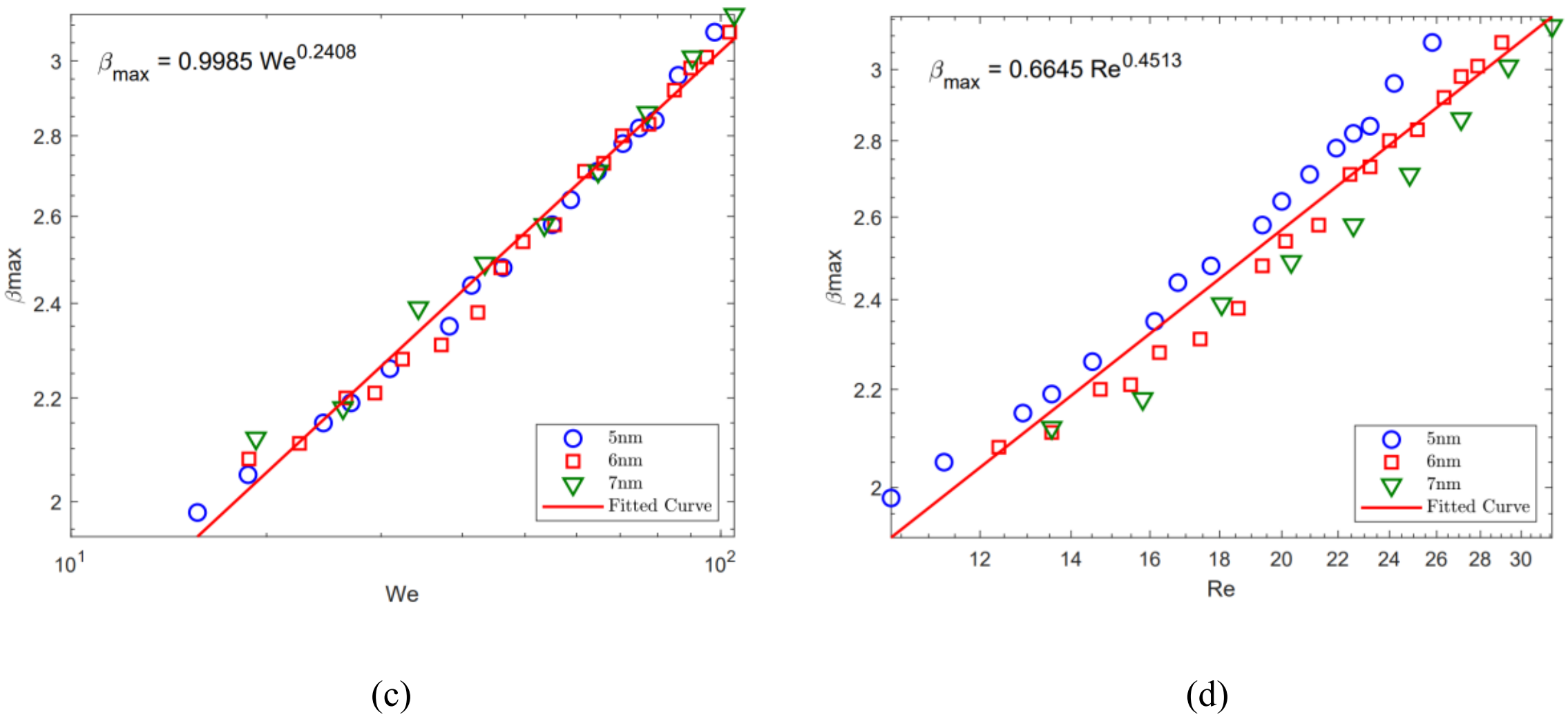

(c) (d)

**Figure 13.** Spreading factor, βmax as a function of We and Re number in (a), (b) low We and Re number regime and (c), (d) high We and Re number regime. The data points that are fitted for the droplet radius 5, 6, and 7 *nm* on a flat superhydrophobic surface (CA ~ 180).

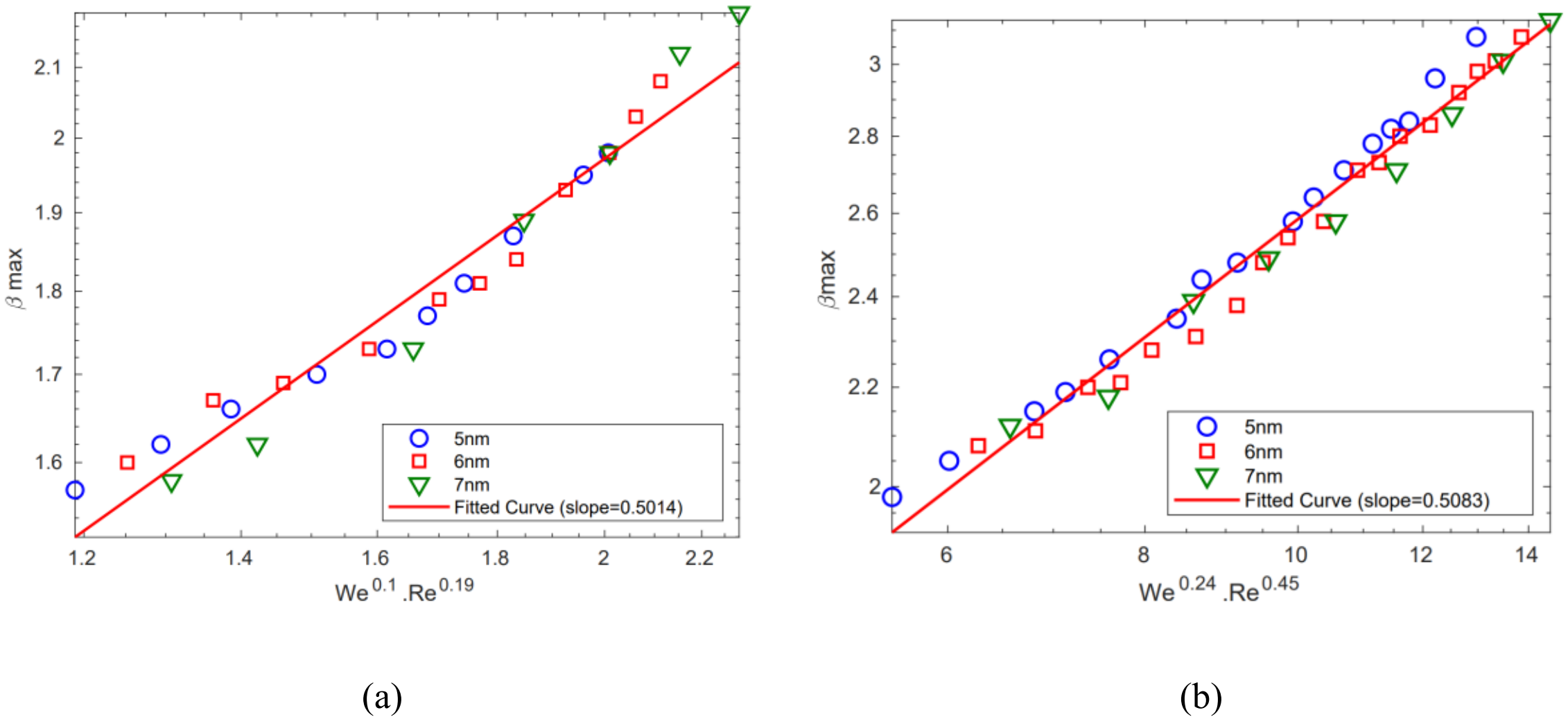

(a) (b)

**Figure 14.** Variation of spreading factor, $\beta_{max}$ with both We and Re number (a) for low We number regime and (b) for high We number regime. Three data sets shown in the graph are for droplets having radii of 5, 6, and 7 *nm* on a flat superhydrophobic surface (CA ~ 180).

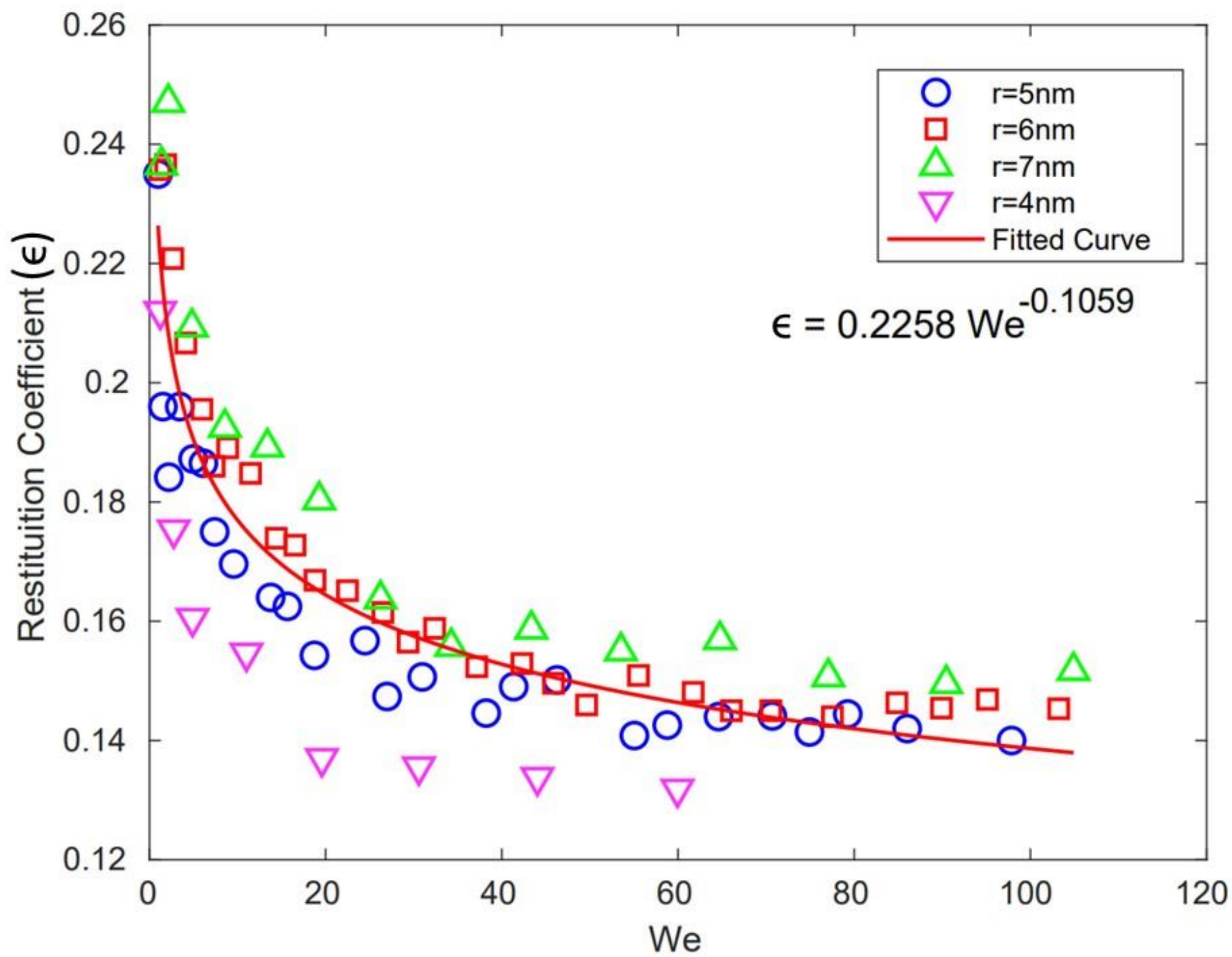

**Figure 15.** Restitution coefficient as a function of we number for droplets having radius 4, 5, 6, and 7 *nm*.

## 4. Conclusion

The study of impinging droplets on a stationary droplet over a solid substrate has gained considerable attention in both industrial and academic research endeavors. Understanding the underlying physics of this phenomenon is crucial for advancing its practical applications. This work studies the collision between a moving droplet and a droplet on a solid surface that is stationary, through molecular dynamics simulation. From the simulation results, the following conclusions can be drawn:

- The process of a moving droplet colliding perpendicularly with an immobile droplet on a solid surface, leading to induced jumping, unfolds in five stages:

impacting, coalescing, spreading, receding, and jumping. Insights into this process reveal its distinctions from the impact of a single droplet on the solid.

- Energy conversion lies at the heart of this phenomenon, with the kinetic energy of the moving droplet and the surface energy of the individual droplets serving as primary energy sources. A fraction of the energy (approximately 4%) is utilized to induce the jumping velocity of the merged droplet, while the rest is dissipated through adhesion work (around 1%) and viscous dissipation (approximately 95%).
- Molecular dynamics (MD) simulations conducted on surfaces with varying roughness and wettability shed light on how these factors influence the phenomenon. Induced velocity increases with surface roughness, droplet size, and surface hydrophobicity. The induced velocity curve shifts upward by almost 10 $m/s$ when roughness is created on the surface. A decrease in contact angle for a flat surface by only 15 degrees from 180 degrees causes the minimum impact velocity for jumping to shift from 50 $m/s$ to 150 $m/s$.
- Governing parameters such as maximum spreading time, spreading factor, and restitution coefficient are derived through power-fitting data points obtained from MD simulations across all scenarios. Subsequent development of modified scaling laws for this phenomenon is based on these parameters. The spreading time varies linearly with $We^{0.31}$ for a droplet impacting another droplet, instead of $We^{0.40}$ for a single droplet impact. Similarly, the restitution coefficient has been modified from $\epsilon \sim We^{-0.341}$ to $\epsilon \sim We^{-0.106}$ and for spreading factor, the scaling laws are modified from $\beta_{max} \sim We^{0.20}$ to $\beta_{max} \sim We^{0.05}Re^{0.10}$ for low We numbers and from $\beta_{max} \sim We^{2/3}Re^{-1/3}$ to $\beta_{max} \sim We^{0.12}Re^{0.23}$ for high We number regime, by considering both We and Re numbers in both regimes.

## CRediT authorship contribution statement

**Ertiza Hossain Shopnil:** Writing – original draft, Visualization, Validation, Methodology, Investigation, Data curation. **Jahid Emon:** Writing – review & editing, Methodology, Investigation, Formal analysis. **Md. Nadeem Azad:** Writing – review & editing, Writing – original draft, Visualization, Validation, Methodology, Investigation, Data curation. **A.K.M. Monjur Morshed:** Validation, Supervision, Software, Resources, Formal analysis, Conceptualization.

## Declaration of competing interest

The authors declare that they have no known competing financial interests or personal relationships that could have appeared to influence the work reported in this paper.

## Data availability

Data will be made available on request.